\documentclass[]{aastex701}
\usepackage{CJK}
\begin{document}
\begin{CJK*}{UTF8}{gbsn}

\title{In-flight Characteristics and Modelling of the Instrumental Background of EP/FXT}

\author[orcid=,gname=Juan, sname=Zhang]{Juan Zhang}
\affiliation{State Key Laboratory of Particle Astrophysics, Institute of High Energy Physics, Chinese Academy of Sciences, Beijing, 100049, China}
\email[]{zhangjuan@ihep.ac.cn}  

\author[orcid=,gname=Yong, sname=Chen]{Yong Chen}
\affiliation{State Key Laboratory of Particle Astrophysics, Institute of High Energy Physics, Chinese Academy of Sciences, Beijing, 100049, China}
\email[]{ychen@ihep.ac.cn} 

\author[orcid=,gname=Shumei, sname=Jia]{Shumei Jia}
\affiliation{State Key Laboratory of Particle Astrophysics, Institute of High Energy Physics, Chinese Academy of Sciences, Beijing, 100049, China}
\email[]{jiasm@ihep.ac.cn}

\author[orcid=,gname=Haisheng, sname=Zhao]{Haisheng Zhao}
\affiliation{State Key Laboratory of Particle Astrophysics, Institute of High Energy Physics, Chinese Academy of Sciences, Beijing, 100049, China}
\email[]{zhaohs@ihep.ac.cn} 

\author[orcid=,gname=Weiwei, sname=Cui]{Weiwei Cui}
\affiliation{State Key Laboratory of Particle Astrophysics, Institute of High Energy Physics, Chinese Academy of Sciences, Beijing, 100049, China}
\email[]{cuiww@ihep.ac.cn}

\author[orcid=,gname=Tianxiang, sname=Chen]{Tianxiang Chen}
\affiliation{State Key Laboratory of Particle Astrophysics, Institute of High Energy Physics, Chinese Academy of Sciences, Beijing, 100049, China}
\email[]{chentx@ihep.ac.cn}

\author[orcid=,gname=Juan, sname=Wang]{Juan Wang}
\affiliation{State Key Laboratory of Particle Astrophysics, Institute of High Energy Physics, Chinese Academy of Sciences, Beijing, 100049, China}
\email[]{wangjuan@ihep.ac.cn}

\author[orcid=,gname=Hao, sname=Wang]{Hao Wang}
\affiliation{State Key Laboratory of Particle Astrophysics, Institute of High Energy Physics, Chinese Academy of Sciences, Beijing, 100049, China}
\email[]{wanghao87@ihep.ac.cn}

\author[orcid=,gname=Jin, sname=Wang]{Jin Wang}
\affiliation{State Key Laboratory of Particle Astrophysics, Institute of High Energy Physics, Chinese Academy of Sciences, Beijing, 100049, China}
\email[]{jinwang@ihep.ac.cn}

\author[orcid=,gname=Chengkui, sname=Li]{Chengkui Li}
\affiliation{State Key Laboratory of Particle Astrophysics, Institute of High Energy Physics, Chinese Academy of Sciences, Beijing, 100049, China}
\email[]{lick@ihep.ac.cn}

\author[orcid=,gname=Xiaofan, sname=Zhao]{Xiaofan Zhao}
\affiliation{State Key Laboratory of Particle Astrophysics, Institute of High Energy Physics, Chinese Academy of Sciences, Beijing, 100049, China}
\email[]{zhaoxf@ihep.ac.cn}

\author[orcid=,gname=Ju, sname=Guan]{Ju Guan}
\affiliation{State Key Laboratory of Particle Astrophysics, Institute of High Energy Physics, Chinese Academy of Sciences, Beijing, 100049, China}
\email[]{jguan@ihep.ac.cn}

\author[orcid=,gname=Dawei, sname=Han]{Dawei Han}
\affiliation{State Key Laboratory of Particle Astrophysics, Institute of High Energy Physics, Chinese Academy of Sciences, Beijing, 100049, China}
\email[]{dwhan@ihep.ac.cn}

\author[orcid=,gname=Jingjing, sname=Xu]{Jingjing Xu}
\affiliation{State Key Laboratory of Particle Astrophysics, Institute of High Energy Physics, Chinese Academy of Sciences, Beijing, 100049, China}
\email[]{xujingjing@ihep.ac.cn}

\author[orcid=,gname=Liming, sname=Song]{Liming Song}
\affiliation{State Key Laboratory of Particle Astrophysics, Institute of High Energy Physics, Chinese Academy of Sciences, Beijing, 100049, China}
\email[]{songlm@ihep.ac.cn}

\author[orcid=,gname=Hua, sname=Feng]{Hua Feng}
\affiliation{State Key Laboratory of Particle Astrophysics, Institute of High Energy Physics, Chinese Academy of Sciences, Beijing, 100049, China}
\email[]{hfeng@ihep.ac.cn}

\author[orcid=,gname=Shuangnan, sname=Zhang]{Shuangnan Zhang}
\affiliation{State Key Laboratory of Particle Astrophysics, Institute of High Energy Physics, Chinese Academy of Sciences, Beijing, 100049, China}
\email[]{zhangsn@ihep.ac.cn}

\author[orcid=,gname=Weimin, sname=Yuan]{Weimin Yuan}
\affiliation{National Astronomical Observatories, Chinese Academy of Sciences, Beijing, 100012, China}
\email[]{wmy@nao.cas.cn}

\begin{abstract}

The in-flight instrumental background of the Follow-up X-ray Telescope (FXT) onboard Einstein Probe (EP) mission is analysed in this work by utilizing observations collected during Performance Verification phase and subsequent dedicated filter wheel closed observations. The instrumental backgrounds of the two FXT modules are consistent with each other, with an average rate of $\sim 4\times10^{-2}$\,counts/s/keV in the 0.5--10\,keV band for each module. The background is nearly uniformly distributed across the detector area, with a minor increase ($<8\%$) observed along rows. The spatial distribution shows significant modulation by the geomagnetic field. The spectral shapes remain unchanged in 0.5--10\,keV at different rates. The long-term temporal variation indicates a periodic change associated with the orbital precession ($\sim 57$ days). The innovative design of FXT full-frame readout mode enables simultaneous recording of events in both the imaging area (IMG) and the frame store area (FSA) of the pnCCD. FSA event rates show a strong linear correlation with the IMG, based on which the IMG instrumental background modelling is established. 

\end{abstract}


\section{Introduction} \label{sec:intro}

Einstein Probe (EP) is an international collaboration mission led by the Chinese Academy of Sciences (CAS) with the European Space Agency (ESA), the Max Planck Institute for Extraterrestrial Physics (MPE) and the Centre National d’Etudes Spatiales (CNES) \citep{2022hxga.book...86Y}. It is dedicated to the time domain of the X-ray astrophysics with the motivation to explore cosmic high-energy transients and monitor variable objects \citep{2016SSRv..202..235Y,2018SSPMA..48c9502Y,2018SPIE10699E..25Y,2022hxga.book...86Y}. At 15:03 Beijing Time on January 9th, 2024, EP was successfully launched into a low-Earth orbit (LEO) with an altitude of $\sim590$\,km and an inclination angle of $29^\circ$. The two scientific payloads onboard are the Wide-field X-ray Telescope (WXT, micro-pore lobster-eye optics + CMOS, 0.5--4\,keV, \citealt{2024ExA....57...10C}) and the Follow-up X-ray Telescope (FXT, Wolter type-I optics + pnCCD, 0.5--10\,keV, \citealt{2025RDTM..tmp...65C}). These instruments are designed to collaborate and complement each other synergistically. WXT fully takes advantage of its wide field-of-view (FOV) of 3600 square degrees ($\sim$1.1\,Sr) and high sensitivity \citep{2017ExA....43..267Z} to capture transients and monitor variable celestial objects. FXT conducts deep follow-up observations to dissect the underlying astrophysical processes by virtue of its larger effective area, better spatial and temporal resolutions \citep{2025RDTM..tmp...65C,2023ExA....55..603C,2023ExA....55..625Y,2025RAA....25a5002Z}.

For space-borne X-ray instruments, precise in-orbit background subtraction is crucial for scientific analysis, particularly when dealing with faint and extended sources. Generally, the background is induced by space environment and the electronic noise. The former can be divided into sky background and instrumental background. Sky background typically refers to direct contributions incident on the sensitive detector through the instrument's FOV or optics, such as the diffuse X-ray emission focused by the Wolter type-I mirror, which directly interacts with and is then detected by the pnCCD in EP/FXT. Instrumental background results from the interaction of space environment particles (in most cases omnidirectional) with the detector directly or with surrounding materials, which generate secondary particles that are then detected by the detector. As estimated in the pre-launch simulation \citep{2022APh...13702668Z}, for EP/FXT the dominant instrumental background components are induced by cosmic rays and the cosmic X-ray background outside the FOV. Electronic noise depends on the detector's characteristics and operational parameters, such as voltage, temperature, and threshold. It is mostly present at low energies, e.g. below 0.2\,keV for FXT, and can be suppressed by selecting suitable operational parameters.

In this study, we examine the characteristics of the in-orbit instrumental background of EP/FXT, focusing on aspects such as the background spectrum and rate, as well as their spatial and temporal distributions. Based on the novel full-frame mode (FF) readout design of FXT, we explore the correlation between the instrumental background and the frame store area (FSA) data to develop an estimation model. This paper begins with an introduction to the EP/FXT configuration in Section \ref{sec::epfxt}. The observations used and the data reduction are detailed in Section \ref{sec::obsdata}. Subsequently, the properties of the instrumental background are analyzed in Sections \ref{sec:dis_on_pnccd}, \ref{sec:rate_and_spc} and \ref{sec:orbit_variation}, followed by an investigation of the correlation and background modelling in Section \ref{sec::bkgmodel}. With this correlation, the long-term variation over one year is explored in Section \ref{sec:long_term_variation}. The final conclusions are summarized in Section \ref{sec::con}.

\section{EP/FXT\label{sec::epfxt}}

\begin{figure}[htb]
    \centering
    \includegraphics[width=7cm]{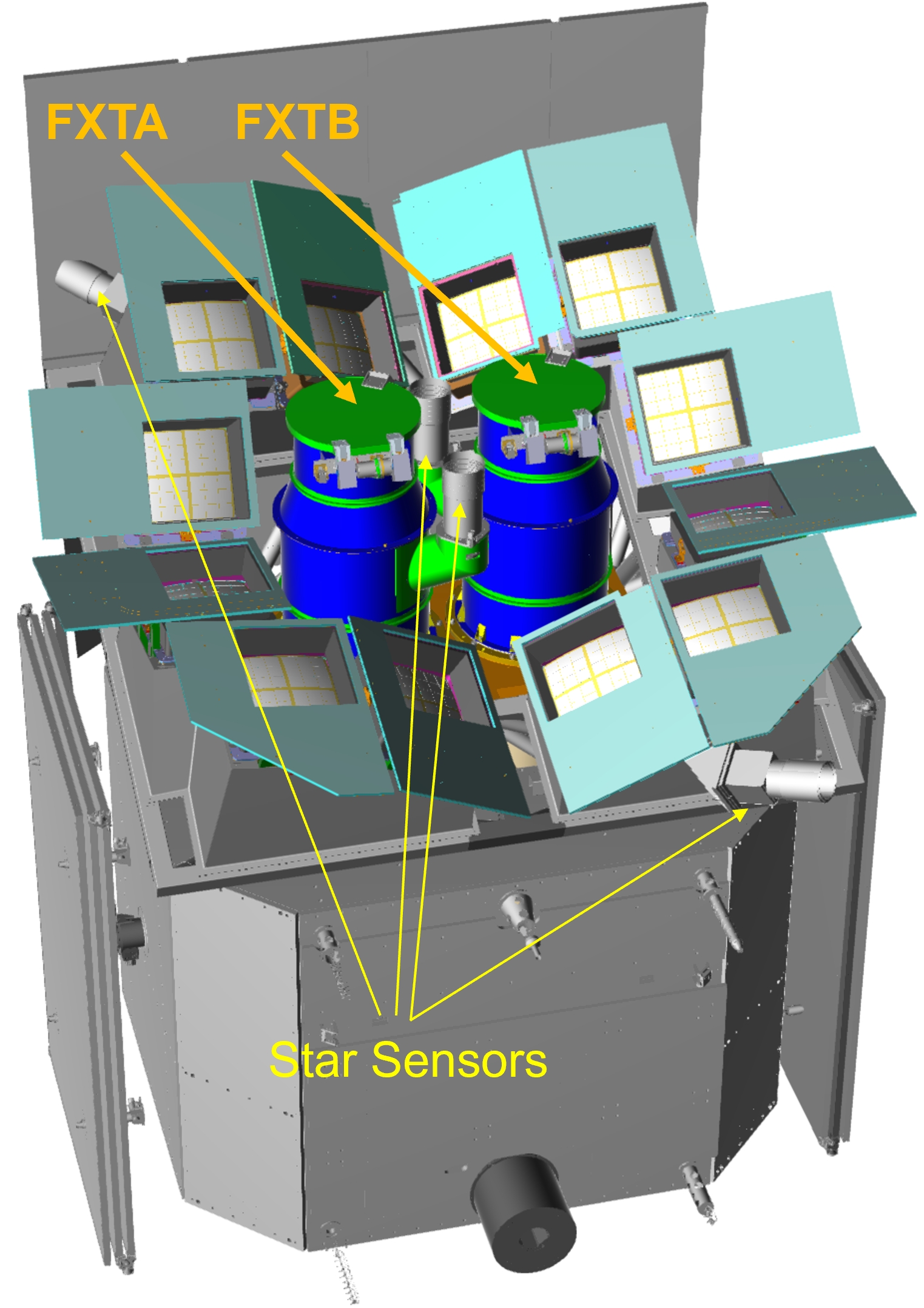}
    \caption{EP structure before the solar panels are deployed.}
    \label{fig::ep}
\end{figure}

\begin{figure}[htb]
    \centering
    \includegraphics[width=1\linewidth]{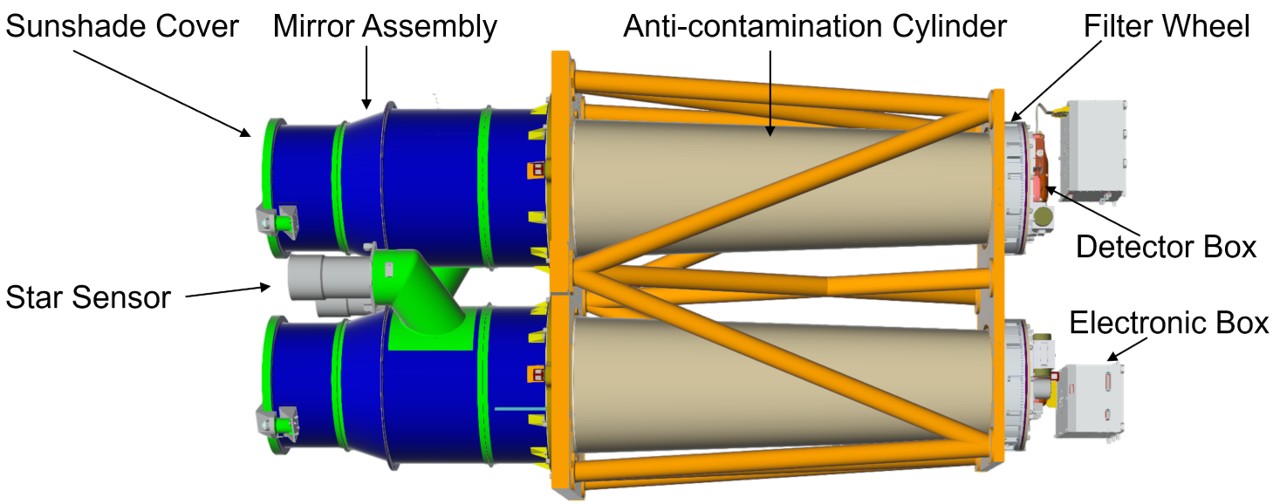}
    \caption{EP/FXT two modules.}
    \label{fig::fxt_configuration}
\end{figure}

EP/FXT consists of two identical co-aligned modules, namely FXTA and FXTB respectively, which are surrounded by the 12 WXT modules. Figure\,\ref{fig::ep} illustrates the complete structure of EP before the solar panels are deployed, and Figure\,\ref{fig::fxt_configuration} depicts the two FXT modules. The sunshade cover at the top of each mirror assembly is made up of a sandwich structure comprising two 0.4\,mm thick carbon fiber face sheets (top and bottom) and a 24.2\,mm thick aluminum honeycomb core in between. It is used to protect the optics, and was opened to a small angle of $\sim 3^\circ$ after launch for outgassing before being completely uncapped for celestial observations. Each FXT module is similar to an individual SRG/eROSITA telescope \citep{2021A&A...647A...1P} due to the fact that FXTB utilizes the flight spare mirror assembly (MA) of SRG/eROSITA provided by MPE as its flight model, FXTA employs an equivalent MA supplied by ESA, and both FXT modules incorporate pnCCDs provided by MPE that are nearly identical to those on SRG/eROSITA with only minor differences in the on-chip filter thickness \citep{2025RDTM..tmp...65C,2023ExA....55..603C,2022APh...13702668Z}. FXTA and FXTB are capable of working independently in different filter wheel positions and pnCCD readout modes. The pnCCDs of these two modules are placed on the focal plane with the readout orientations orthogonal to each other.  

\subsection{Filter Wheel}

\begin{figure}[htb]
    \centering    
    \includegraphics[width=\linewidth]{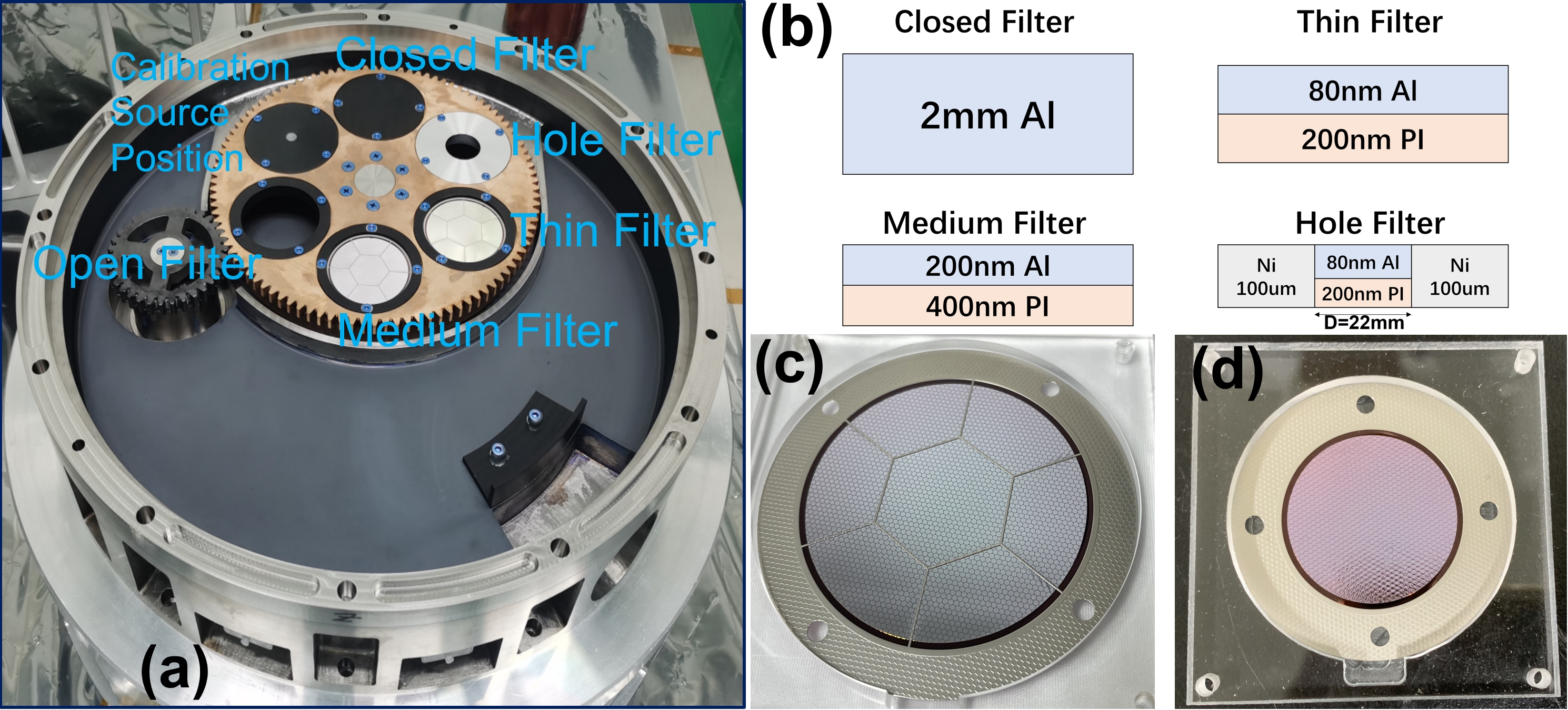}
    \caption{EP/FXT filter wheel and the filters. (a) The six positions on the filter wheel. (b) The schematic design diagrams of the closed, thin, medium and hole filters. (c) The large hexagonal frame structures and the tiny hexagonal grids in the thin/medium filter. (d) The small aperture hole filter, surrounded by the nickle annulus.}
    \label{fig::filter}
\end{figure}

The filter wheel of EP/FXT was designed and constructed by the Institute of High Energy Physics, CAS (IHEP) \citep{2023ExA....55..603C,2023OptEn..62d4103C}. There are six operational positions: the open, thin, medium and hole filters, the closed filter and the calibration source position, as shown in the picture of Figure\,\ref{fig::filter}(a). Located at about 7\,cm above the top surface of the pnCCD, these filters and positions enable versatile observational capabilities. The calibration source utilizes $^{55}$Fe with aluminum target, which emits characteristic X-rays at Mn-K$_\alpha$ (5.895 keV), Mn-K$_\beta$ (6.490 keV) and Al-K$_\alpha$ (1.49 keV) photons for in-orbit energy calibration. The open position is currently unavailable to the scientific community, considering the potential risk of encountering micrometeoroids. The schematic design diagrams of the other four types of filters are illustrated in Figure\ref{fig::filter}(b). The closed filter, comprising a 2\,mm-thick aluminum (Al) layer, facilitates instrumental background measurements by blocking external X-rays.  The thin filter employs 80\,nm Al + 200\,nm Polyimide (PI). The medium filter differs in thickness, which is 200\,nm Al + 400\,nm PI in contrast. The hole filter layers resemble those of the thin filter, but they feature a smaller aperture with a diameter of 22\,mm, encased within the surrounding nickel (Ni) annulus, as visualized in Figure\,\ref{fig::filter}(d). There are tiny hexagonal grids to form the filters, which is PI for the thin and hole filters, and Ni for medium filter. In addition, large hexagonal nickel frames are used both in the thin and medium filters for structure supporting, which is apparently seen in Figure\ref{fig::filter}(c).

\subsection{pnCCD and Readout Modes\label{sec::pnccd_modes}}

The pnCCDs of FXT were provided by MPE \citep{2022hxga.book...86Y}. They are similar to those used in SRG/eROSITA. There are $384\times384$ pixels both in the image area (IMG) and in the FSA. Each pixel size is 75\,$\mu$m$\times$75\,$\mu$m in IMG and 51\,$\mu$m$\times$75\,$\mu$m in FSA. The difference is that the thickness of the on-chip aluminum filter is 90 nm, followed by 30\,nm Si$_3$N$_4$ and 20\,nm SiO$_2$ above the 450\,$\mu$m silicon depletion layer \citep{2022APh...13702668Z}. Each pnCCD is placed inside an aluminum alloy detector box, which is nested into a $\sim$3\,cm oxygen-free copper box for shielding the cosmic ray particles and photon backgrounds \citep{2023ExA....55..603C}. 

IHEP designed three readout modes for scientific observations \citep{2023ExA....55..603C}, i.e. the full frame mode (FF), the partial windowed mode (PW) and the timing mode (TM). In FF, the frame is readout every 50\,ms, where the integration time of the IMG is 49.88485\,ms and the 0.11515\,ms corresponds to the fast transfer time from IMG to FSA. There is a delay time of 4.28485\,ms before read out. The total readout time of the rows lasts 9.1168\,ms. FXT has an innovative electronic readout design that the FSA is also integrated for a time of 25\,ms and read out during the IMG integration time in FF. In PW, a smaller region of the pnCCD, currently 128 columns $\times$ 61 rows, is read out per 2.2ms. In TM, all the IMG rows in the central 128 columns are shifted and readout with a speed of 23.68\,$\mu$s/row; after this, an additional 342.08\,$\mu$s time is used to reset the electronics per frame. The FF and PW has two dimensional imaging ability. While in TM the 2D spatial information is projected to 1D. Therefore the orthogonal readout direction arrangement of the two FXT modules could help source resolution and localization when FXTA and FXTB both work in TM modes.

\section{Data and Reduction\label{sec::obsdata}}

As soon as EP was launched into orbit, the sunshade covers at the top of FXTA and FXTB MAs immediately opened a very small angle for outgassing. This $3^\circ$ opening was maintained from the beginning of the performance verification and calibration (PV-CAL) phase until the end of February 2024, after which they were totally uncapped for celestial observations. The data during this period are labeled as sunshade closed data (SCD) in this paper. There were also dedicated observations for FXTA and FXTB separately setting the filter wheel to closed filter position after PV-CAL phase. The filter wheel closed data (FWC), as well as the SCD, are used in this work to investigate the in-orbit instrumental background of EP/FXT. All observations used are listed \href{https://ihepbox.ihep.ac.cn/ihepbox/index.php/s/Q5WD8pD7ar75tfN}{online}. 
It is worth noting that all of the selected data are observed in FF mode. The backgrounds in PW and TM modes are not included in this work due to insufficient observation statistics\footnote{The instrumental backgrounds in PW and TM modes could not be obtained by simply scaled from those of FF mode.}.

The observation data are reduced using \href{http://epfxt.ihep.ac.cn/analysis}{FXT Data Analysis Software V1.10 and FXT calibration database (CALDB) V1.10}. The pipeline \textit{fxtchain} with default good time interval (GTI) selection criteria, ``ELV\textgreater{}5\&\&COR\textgreater{}6\&\&SAA==0\&\&DYE\textunderscore{}ELV\textgreater{}30'', and clean event selections (e.g. grade 0-12 and status==b0), are used. 

The effective exposures accumulated are listed in Table\,\ref{tab:particel_bkg}, which is about 537\,ks for FXTA SCD in thin filter and 609\,ks for FXTB SCD in thin filter. There is also a set of SCD of FXTB in the hole filter with an effective exposure of 141\,ks. All of the SCD were observed in February, 2024. The FXTA FWC were obtained in July, 2024 with an effective exposure of $\sim$ 61\,ks,  and the FXTB FWC in January, 2025 with an effective exposure of $\sim$67\,ks.

\begin{table}[htb]
\caption{The instrumental background observation logs}
\label{tab:particel_bkg}
\centering
\begin{tabular}{cccccll}
\hline
\hline
     & module & filter & exposure& rate@0.5--10keV\\
     &        &   & ks  & $\times 10^{-1}$ counts/s \\
\hline
 
SCD  & FXTA& thin  & 536.8 &  3.61 $\pm$ 0.01   \\
  & FXTB& thin  & 609.0 &  3.71 $\pm$ 0.01   \\
  & FXTB& hole  & 141.3 &  3.96 $\pm$ 0.02   \\
\hline
FWC   & FXTA& closed  &  61.2 &  3.72 $\pm$ 0.02   \\
  & FXTB& closed  &  67.1 &  3.83 $\pm$ 0.02   \\
 \hline
 \hline
\end{tabular}
\end{table}

\section{Distribution on pnCCD\label{sec:dis_on_pnccd}}

\begin{figure}[htb]
\centering
\includegraphics[width=\linewidth]{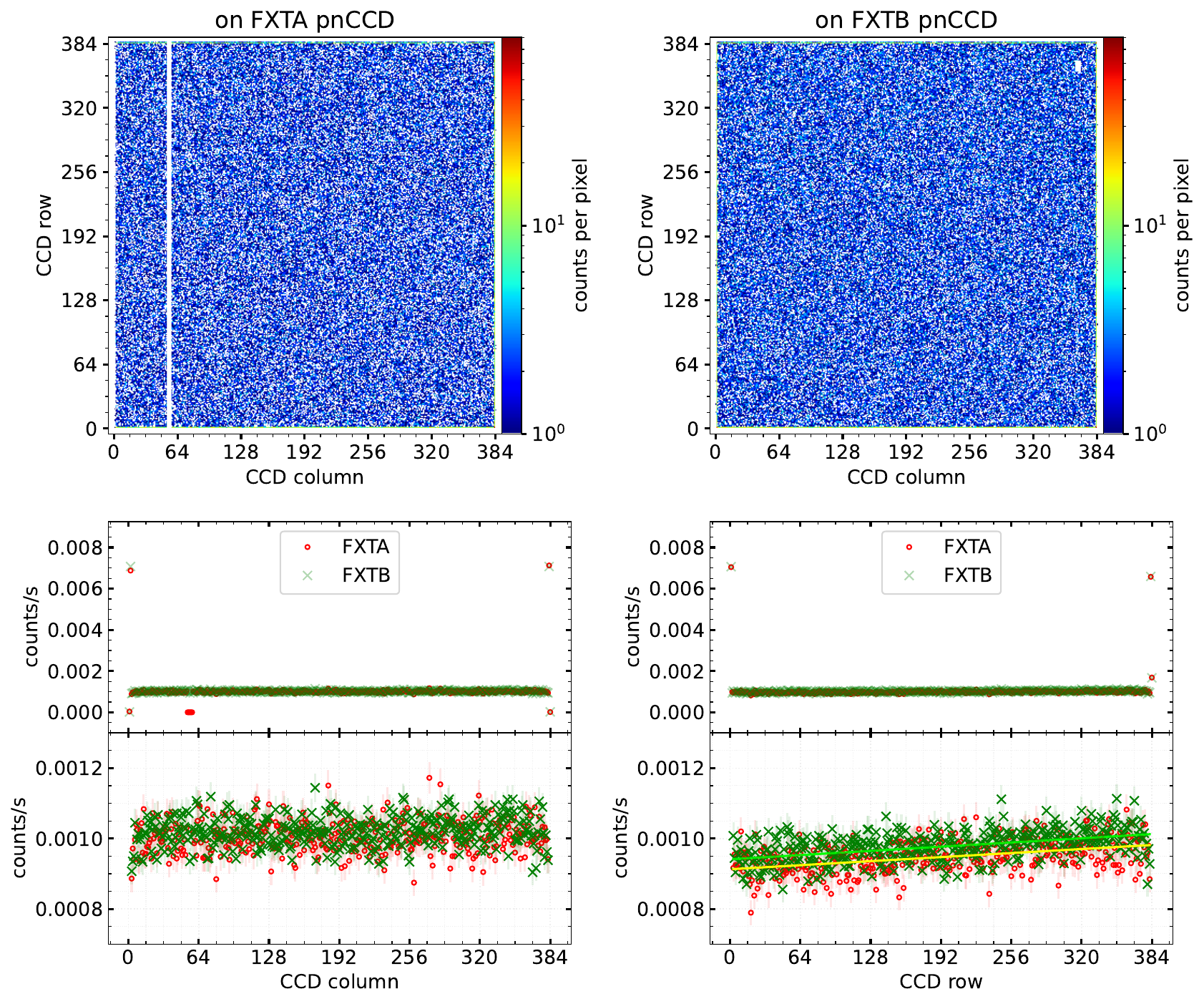}
\caption{Top: The distribution of instrumental background events in the energy range of 0.5 to 10\,keV on the IMG pixels of the pnCCD for FXTA and FXTB, respectively. Bottom: The 1-D projection distribution to the columns and rows, FXTA in red circles and FXTB in green crosses.}
\label{fig::ccd_dis}
\end{figure}

The distributions on the pnCCD area of clean events in the energy band of 0.5 to 10\,keV are presented in the top panels of Figure\,\ref{fig::ccd_dis} for FXTA and FXTB, respectively. The CCD column and row range from 1 to 384. The absence of events at five columns centered on Column 56 and around the pixel (Column 327, ROW 128) on FXTA, as well as around the pixel (Column 365, ROW 358) on FXTB, are attributed to their bad pixel characteristics, which are removed during data reduction. As shown in the lower panels, rates on the outermost two columns and two rows deviate from the average values, which is resulted from the edge effect caused by the grade calculation algorithm using adjacent $3\times3$ pixels. Therefore the inner pixels, $\sim 380\times375$ pixels for FXTA, and $\sim 380\times380$ for FXTB, are retained for the background analysis in the following sections. The projections on the columns and rows indicate that the distribution on pnCCD columns are uniform, and that there is a slight increase of $\sim$7.6\% from Row 3 to Row 382, which might due to the longer exposure time before being read out for farther row pixels. The slope of this increase along rows, fitted by intercept*( 1+slope*row), is (2.0$\pm$0.2)$\times 10^{-4}$. This distribution is similar to the minimum ionizing particles distribution measured by SRG/eROSITA \citep{2021SPIE11444E..1OF}.

\section{Rate and Spectrum\label{sec:rate_and_spc}}

\begin{figure}[htb]
\centering
\includegraphics[width=\linewidth]{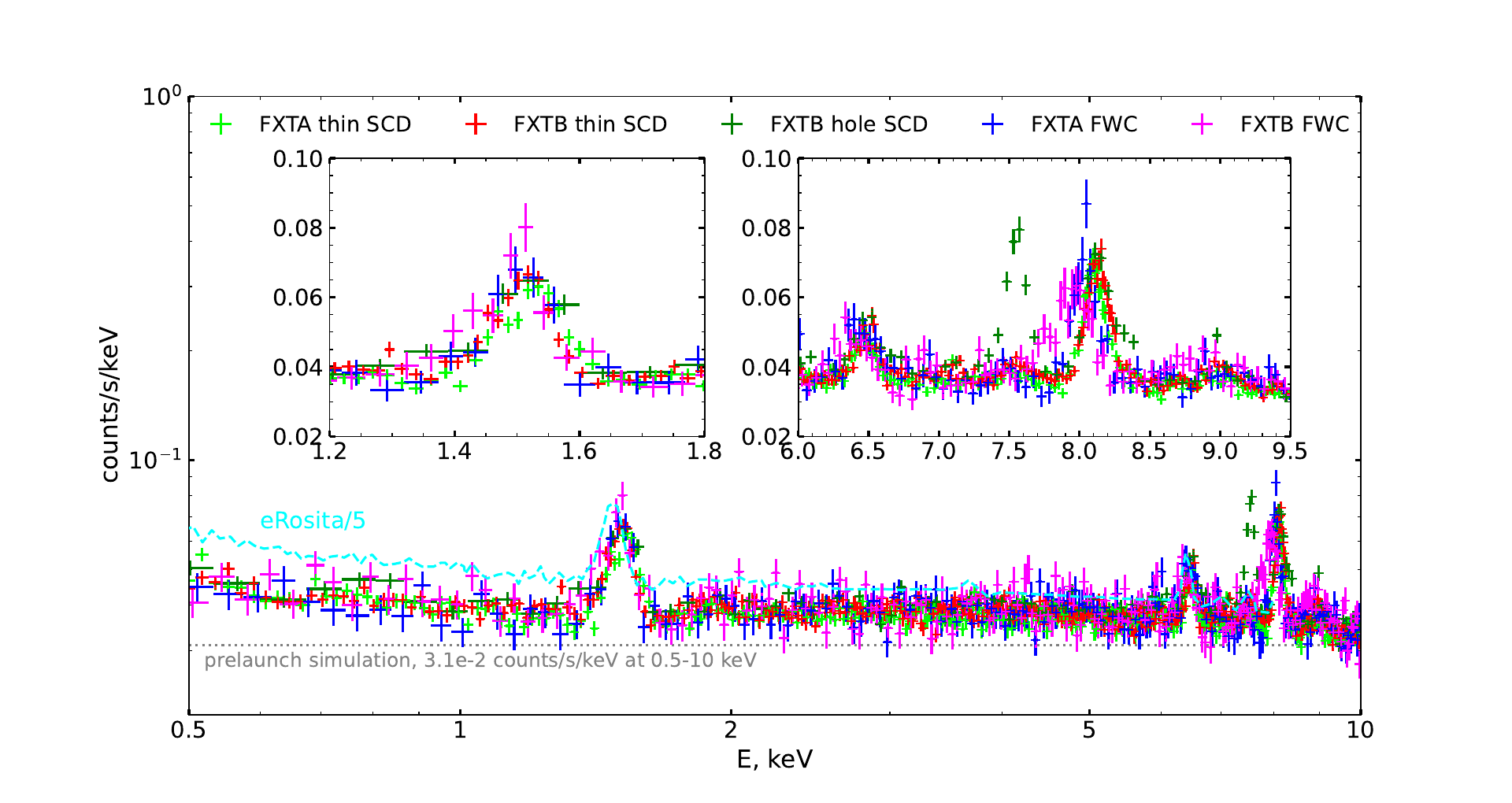}
\caption{The instrumental background spectra of FXTA and FXTB using SCD and FWC data. Also plotted is the pre-launch estimate instrumental background level of one FXT module (in dotted gray), and the SRG/eROSITA instrumental background level derived from \cite{2021A&A...647A...1P} which is rescaled by a factor of $1/5$ for comparison in this plot.}
\label{fig:particle_bkg_spc}
\end{figure}

The instrumental background rate in the 0.5--10\,keV band of the SCD and FWC are also listed in Table\,\ref{tab:particel_bkg}, classified according to the modules and filter wheel positions. The corresponding spectrum of each catalog is plotted in Figure\,\ref{fig:particle_bkg_spc}. The rate values show the subtle nuance among different cases. The largest difference, i.e. between the FXTA thin filter SCD and the FXTB hole filter SCD, are less that 10\%. And the differences between FXTA and FXTB in the same filter position is $\sim$ 3\%. So is the difference for each module between in thin and closed filters. Considering the number of pixels retained after data reduction, the rate and spectral results show that the instrumental background of FXTA and FXTB are consistent for the same filter position. The increased rate in the hole filter position originates from the additional Ni-K$_\alpha$ emissions which is evidently illustrated as the line spectrum structure around 7.5\,keV in Figure\,\ref{fig:particle_bkg_spc}. 

The in-orbit instrumental background rate of one EP/FXT module is in a magnitude of $\sim$ 4$\times10^{-2}$\,counts/s/keV in the 0.5--10\,keV band. Compared with this level, the FXT pre-launch simulation \citep{2022APh...13702668Z}, $\sim$ 3.1$\times10^{-2}$\,counts/s/keV, underestimated the in-orbit measurement by $\sim23\%$. This under estimate may be related to the variable space environment, along with the ideal situations of instrumental and data analysis effects considered in the simulation. The in-orbit instrumental background rate of a single EP/FXT module is approximately 5 times lower than that of an individual SRG/eROSITA module \citep{2021A&A...647A...1P}, which operates in the halo orbit around L2. Since the telescopes of FXT and SRG/eROSITA are similar, the difference between these two instrumental backgrounds is mainly attributed to the space environment in each individual orbit. Considering the background spectrum of SRG/eROSITA was measured around solar minimum, while EP/FXT data has been taken closer to solar maximum, the eROSITA non-X ray background in 2024 and 2025 would be likely at least a factor of 2 lower\footnote{\href{https://iachec.org/wp-content/presentations/2023/Perinati_IACHEC2023.pdf}{``simulation study of the eROSITA NXB'' on 15th IACHEC meeting}} \citep{2021SPIE11444E..1OF} than reported in \citet{2021A&A...647A...1P}. Thus a factor of $\sim$ 2--3 difference between EP/FXT and SRG/eROSITA is anticipated if they were operating under the same solar cycle.

As shown in Figure\,\ref{fig:particle_bkg_spc}, the fluorescence lines at Al-K$_\alpha$ (1.49\,keV), Fe-K$_\alpha$ (6.40\,keV), Cu-K$_\alpha$ (8.05\,keV) and Cu-K$_\beta$ (8.91\,keV) are remarkable in the FXT measured background spectra. These lines come from the surrounded materials around the pnCCD. It is worth noting that the Al-K$_\alpha$ line intensity does not show noticeable enhancement from the thin filter position in SCD to the closed filter position in FWC. This proves that the aluminum constituted the closed filter is not the primary source of the Al-K$_\alpha$ line. Instead, they mainly originate from the aluminum alloy in the detector box and the closed filter only accounts for a minor contribution\footnote{from the SRG/eROSITA simulation analysis in the collaboration conference communication.}. The 6.40\,keV Fe-K$_\alpha$ lines result from the impurities of the beryllium, which was placed under the pnCCD in the detector box for grade shielding. The line at Ni-K$_\alpha$ (7.47\,keV) manifested in the spectrum in hole filter comes from the composition material nickel that was used to constrain the entrance window of the hole filter. Thus it is not significant in the spectra under thin and closed filters. Except for the Ni-K$_\alpha$ line, the spectrum under the hole filter is consistent with spectra in other filter positions. The fluorescence lines of Cu originate from the oxygen-free copper of the detector box. The noticeable displacement of Cu-K$_\alpha$, as well as Fe-K$_\alpha$, in spectra of different epoch implies the temporal change of channel-energy relationship, which is beyond the scope of this work and will be  updated in future CALDB versions and discussed in the forthcoming FXT calibration papers.

\section{Orbital Modulation\label{sec:orbit_variation}}

\begin{figure}[htb]
\centering
\includegraphics[width=\linewidth]{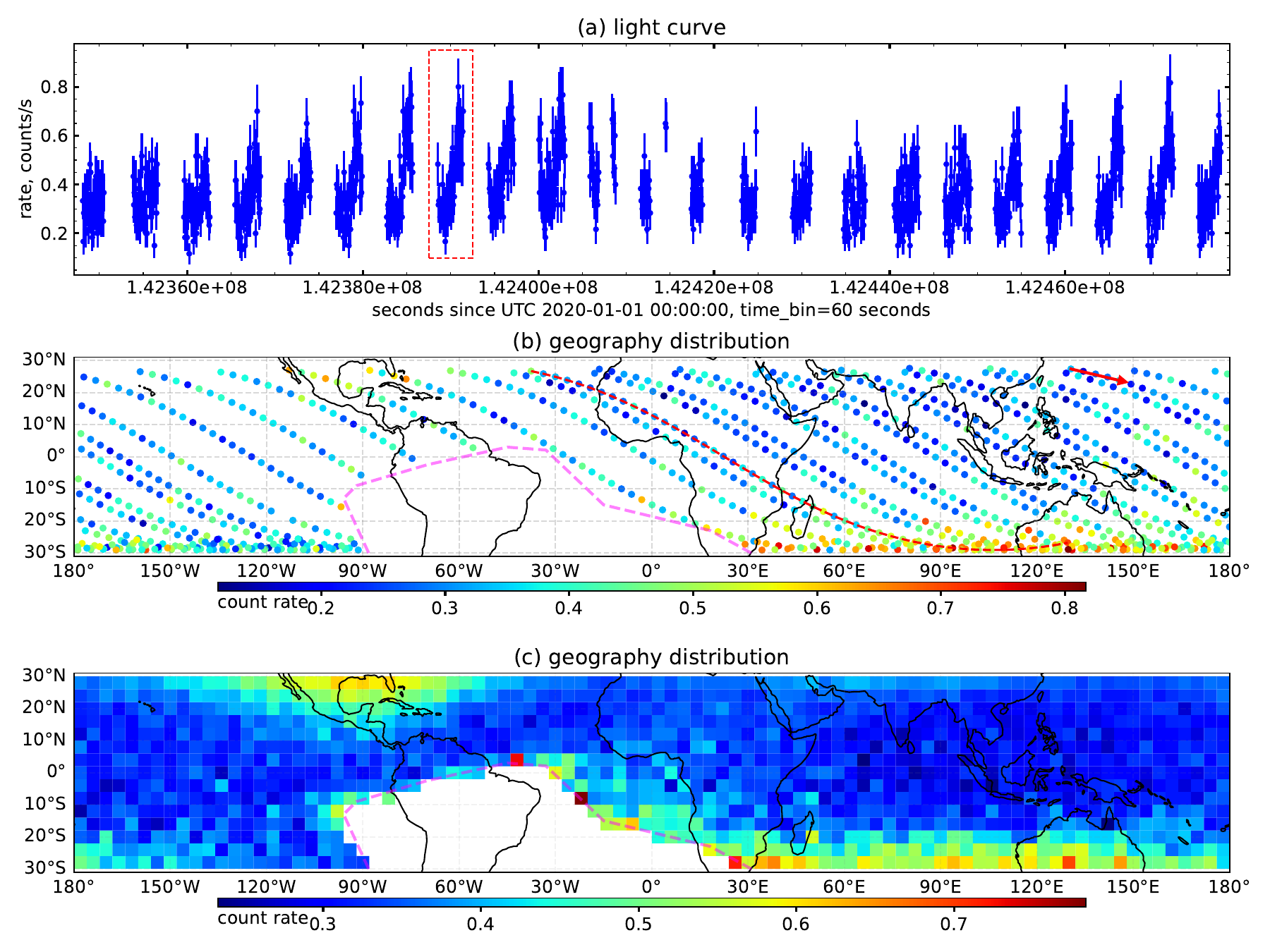}
\caption{The instrumental background variation with time and geographical location.
(a) The light curve in 0.5--10\,keV of one FXTA FWC observation. (b) The corresponding geography distribution of this light curve. (c) The geography distribution of all SCD in thin filter and FWC instrumental background observations. The color bars in Panels (b) and (c) are labeled in units of counts/s, and the SAA region utilized for EP/FXT is delineated by the pink dashed line. Please refer to the text for more detailed descriptions.}
\label{fig:particle_bkg_geo}
\end{figure}

The instrumental background of FXT exhibits orbital modulation of the geomagnetic field, as plotted in Figure\,\ref{fig:particle_bkg_geo}. The top panel plots the light curve of the instrumental background in 0.5--10\,keV by taking the FXTA FWC for instance. The data corresponds to the observation ID (OBSID) 13600006505 and started on July 5th, 2024. The gaps in the light curve were caused by the result of GTI selection, which discarded the periods when the satellite passed through the South Atlantic Anomaly (SAA) and the time intervals with small Earth elevation angles. As illustrated in this panel, for this observation the short term instrumental background fluctuations exhibit a dynamic range of $\sim$ 2--8 times within a single orbit. 

The corresponding geographical longitude and latitude position of each point in the light curve is dotted on the geospatial diagram in the middle panel, where the start of the light curve is marked with the red arrow on the top right, and the orbit enclosed with the dashed red rectangle in Panel (a) is illustrated using the red dashed line in Panel (b). It is clearly seen that the high background rate happens at the intense space radiation environment near 30$^\circ$S latitude.

The entire instrumental background rate distribution on the orbit is plotted in the bottom panel of Figure\,\ref{fig:particle_bkg_geo}, where all the SCD in thin filters and the FWC data are mapped onto the geographical coordinate grids. This panel displays the distribution with enhanced clarity. Apart from the edge of the SAA region, the rate increases towards the north geomagnetic pole direction, i.e. the region near (30$^\circ$N, 90$^\circ$W), and around the mid-latitudes in the south. This pattern is directly correlated with the geomagnetic field, the cut-off rigidity \citep{2009AdSpR..44.1107S, 2021AdSpR..67.2231G} and the LEO proton and electron flux distributions \citep{2014AdSpR..53..233K}. 

\begin{figure}[htb]
    \centering
    \includegraphics[width=\linewidth]{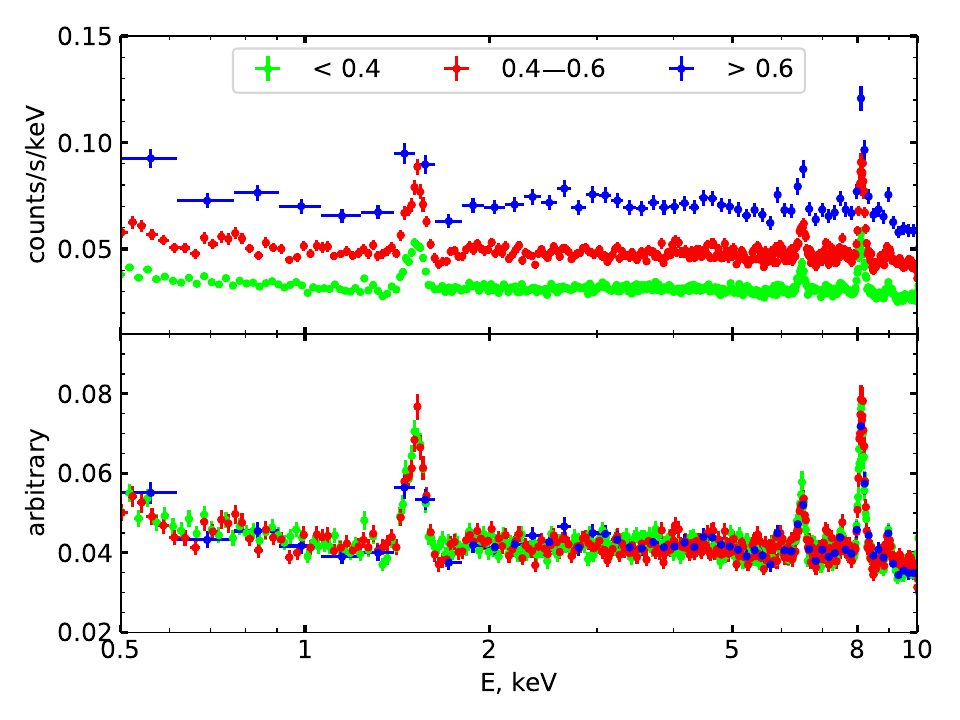}
    \caption{The instrumental background spectra at different rates.}
    \label{fig:spc_shape}
\end{figure}

To examine the spectral variations at varying intensities, we classify the instrumental background events of the FXTA FWC observation into three groups based on the count rate value: rate\textless0.4\,counts/s, rate$\in$[0.4,0.6]\,counts/s and rate\textgreater{}0.6\,counts/s. The spectra corresponding to these three groups are depicted  in the top panel of Figure\,\ref{fig:spc_shape}. The normalization was adjusted to facilitate a comparison among the spectral shapes in the lower panel. It could be concluded that the instrumental background spectral shape in the 0.5--10\,keV energy band remains consistent across different count rates, i.e. at varying  geomagnetic fields. Similar features have been observed in the silicon semiconductor detector of LE telescope aboard the Insight-HXMT mission \citep{2020JHEAp..27...24L}. 
The analysed spectra in Figure\,\ref{fig:particle_bkg_spc} also demonstrate that  the instrumental background retains similar morphology and flux levels within one year period, showing no significant variation.


\section{Particle Background Modelling\label{sec::bkgmodel}}

\begin{figure}[htb]
    \centering
    \includegraphics[width=\linewidth]{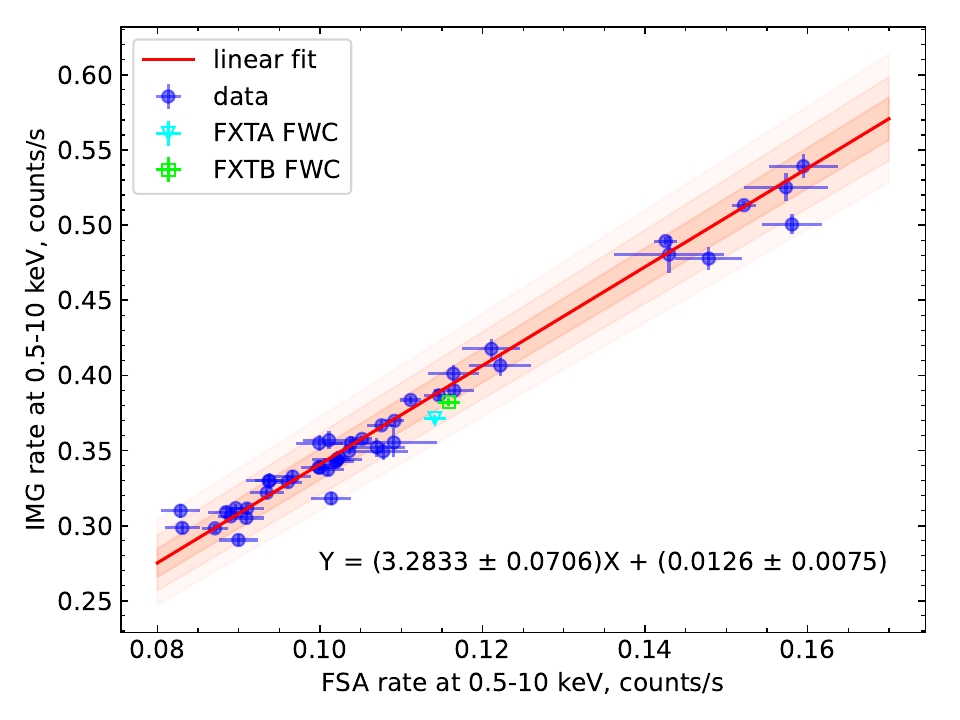}
    \caption{Correlation between the FSA rate and the IMG rate (blue points) in different geography grids. This relation could be well fitted by a linear model (the red line). Also plotted are the observed FSA and IMG rates in the FXTA (cyan triangle) and FXTB (lime square) FWC observations. The shadows indicates the 1$\sigma$, 2$\sigma$ and 3$\sigma$ ranges of the fitted model.}   
    \label{fig:fsa_model}
\end{figure}

The innovative FF readout mode of FXT enables simultaneous integration and readout of FSA during the IMG integration time per frame. Since FSA is located outside the FOV, the recorded events can serve as a real-time indicator of both the space environment radiation and the instrumental background of IMG.  The events in FSA are reduced using the same procedure and selection criteria as those of IMG introduced in Section\,\ref{sec::obsdata}. The correlation between the FSA rate\footnote{It is important to note that the exposure time of FSA events obtained in this procedure is twice the actual value. This discrepancy arises because the cycle time of the IMG (50\,ms) is used other than the true exposure time of the FSA (25 ms) in one frame. Since the FSA rate in this study is treated as an independent variable and absolute values are not employed directly, we retain the results derived from the procedure to represent the FSA count rate.} and the IMG rate is illustrated in Figure\,\ref{fig:fsa_model}. The data points in blue were derived from Figure\,\ref{fig:particle_bkg_geo}(c) by binning the geographic latitudes and longitudes into $10^{\circ}\times45^{\circ}$ grids to reduce statistical uncertainties. The data were fitted using \textit{scipy.odr}, with both IMG and FSA measurement uncertainties incorporated into the model. The linear fit demonstrated excellent agreement with the data, achieving a 1$\sigma$ uncertainty of $\sim$ 3\%. Consistent fitting results were obtained when analyzing the FSA-IMG rates for individual observations; however, there exhibited a larger 1$\sigma$ uncertainty of $\sim$ 12\% due to the limited statistics of the data. The observed FSA and IMG rates for the FWC observations of FXTA and FXTB are marked with the cyan triangle and the lime square, respectively, for comparison. Notably, the predicted IMG rate for the FXTA FWC observation was 4.3\% higher than the observed data, while the prediction for FXTB FWC was 2.8\% higher. The shaded regions represent the 1$\sigma$, 2$\sigma$, and 3$\sigma$ uncertainty ranges of the fitted model, providing insight into the confidence intervals of the linear relationship.

\begin{figure}[tb]
    \centering
    \includegraphics[width=1\linewidth]{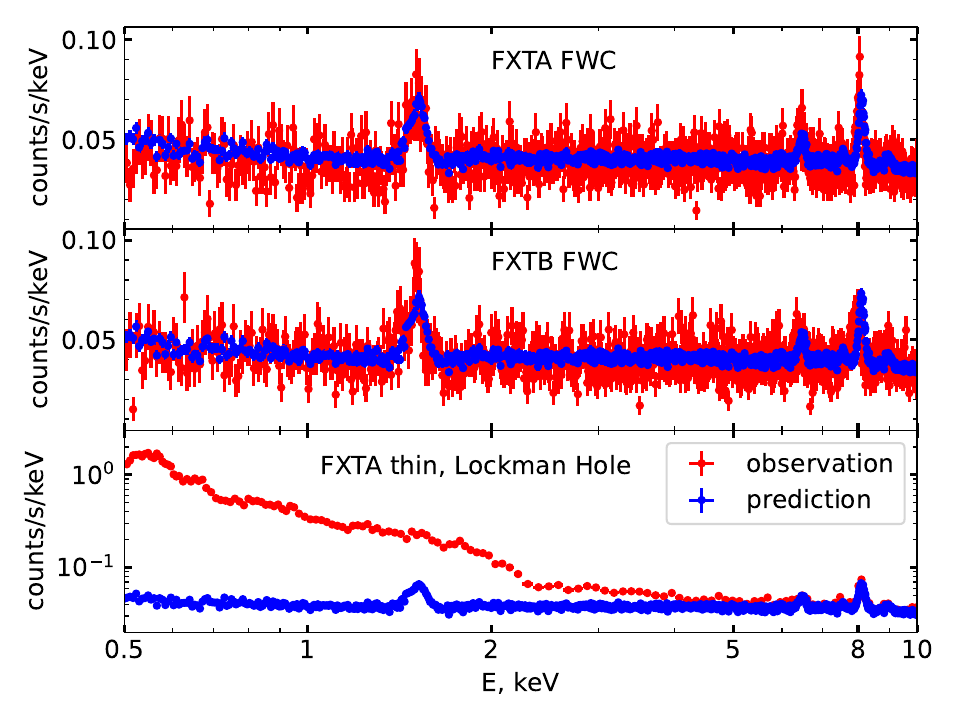}
    \caption{Validation of the instrumental background model. Red points represent observed spectra, and blue points denote predicted ones. The observed spectrum of the Lockman Hole is derived from the source-free region and scaled to the same number of effective pixels as those retained in the instrumental background observations.
    \label{fig:bkgmodel_validation}}
\end{figure}

This linear model was employed to develop the FF IMG instrumental background estimate tool, \textit{fxtbkggen}, which has been available since FXT CALDB Version 1.20. This tool was validated by taking the FWC and the Lockman Hole observations for instances, as illustrated in Figure\,\ref{fig:bkgmodel_validation}. The instrumental background rate was initially predicted based on the linear model and the observation rate of FSA. Since the background shape keeps unchanged across  different rates (evidenced in Figure\,\ref{fig:spc_shape}), all instrumental observations are combined to accumulate a spectral template. The background spectrum for a specific OBSID is obtained by scaling the predicted rate in this observation with that of the template. It shows that the tool could reproduce the instrumental background spectra very well for the FWC observations, and the prediction is compatible with the observed spectrum of the Lockman Hole observation at high energies, where the instrumental background dominates. The estimate error in each channel is calculated through the propagation of the model and the errors of FSA rate and the template.  

\section{Long-term Variation\label{sec:long_term_variation}}

\begin{figure}[htb]
    \centering
    \includegraphics[width=1\linewidth]{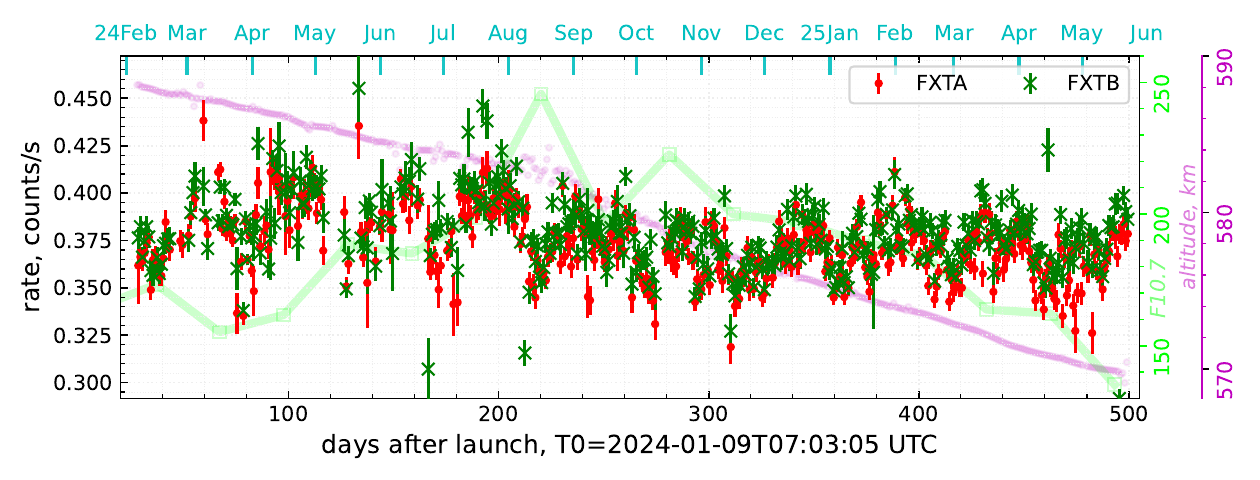}
    \caption{The long term light curves of the derived IMG instrumental background rates, FXTA in red dots and FXTB in green crosses. Also plotted are the 10.7\,cm radio flux (in lime square) from the \href{https://www.swpc.noaa.gov/products/solar-cycle-progression}{Space Weather Prediction Center} and the daily averaged altitude of EP (in magenta circle). The monthly mean F10.7 values are plotted at the midpoint of each month for reference.}
    \label{fig:fsa_longterm_lc}
\end{figure}

To explore the long term temporal variation in the IMG instrumental background rate, we analysed the daily averaged FSA rate from EP launch to the end of May 2025. Utilizing the linear model in Section\,\ref{sec::bkgmodel}, the long-term light curve of the IMG rates in the 0.5--10\,keV energy range was derived, as shown in Figure\,\ref{fig:fsa_longterm_lc}. The amplitude of this long-term variability is approximately $<20\%$. It can be seen that a period of $\sim$ 50 days is implied. This period value aligns with the predicted precession period of the EP orbit ($\sim$ 57 days). 
 
Figure\,\ref{fig:fsa_longterm_lc} also indicates that the daily IMG instrumental background rate during the initial $\sim$ 200 days appears higher than in subsequent epochs. This decrease might be attributed to the fact that in 2024 the Sun was evolving towards its maximum of activity, expected around mid 2025, then the shielding of cosmic rays of the solar potential below a few GeV since launch became more and more efficient. Therefore, the instrumental background rate would exhibit an anti-correlation with the solar activity from a long-term perspective, as viewed by XMM-Newton/EPIC-pn \citep{2020ApJ...891...13B} and Chandra/ACIS \citep{2021A&A...655A.116S,2022SPIE12181E..2EG} based on data accumulated over decade years. As also plotted in Figure\,\ref{fig:fsa_longterm_lc} is the monthly averaged 10.7\,cm radio flux, which serves as a proxy for solar activity. Currently, the particle-induced data collected over one year on FXT have not significantly demonstrated this correlation due to the fluctuation of the solar activity. Further long-term data accumulation is required to analyze the impact of the solar activity on FXT instrumental background. 

As the reduction in orbit height also leads to the decrease in instrumental background rate, due to the shielding effect of the Earth magnetic field against low energy primary cosmic rays, the daily averaged EP altitude is additionally dotted on Figure\,\ref{fig:fsa_longterm_lc} for comparison. Till the end of May 2025, the altitude has degraded approximately 20\,km since launch. And the secular instrumental background rate reduced with time at a speed of $(5.44\pm0.14)\times 10^{-5}$ counts/s per day. The orbit decay could partially account for the secular reduction.

\section{Summary and Conclusion\label{sec::con}}

Based on the PV phase data before the sun-shade covers were completely uncapped and the dedicated filter wheel closed observations during the nominal scientific observation epoch, the properties of the in-orbit instrumental background of EP/FXT were comprehensively investigated in this work. The background levels were found to be consistent between the two FXT modules, each in the magnitude of $4\times10^{-2}$\,counts/s/keV in the energy band of 0.5--10\,keV, which is about 5 times lower than that of a single SRG/eROSITA telescope module. This in-flight magnitude indicates that the pre-launch simulation underestimated by $\sim20\%$. Given the simplification utilized in the detection and data analysis processes in the simulation, and the variable space environment, it was concluded that the pre-launch simulation is reasonable and reliable. 

The in-orbit instrumental background shows nearly uniform distribution across pnCCD pixels except the outermost two columns and two rows resulting from the edging effect of the grade calculation with adjacent $3\times3$ pixels. The minor increase observed along the rows is within $\sim8\%$. The spatial distribution of the instrumental background rate exhibits modulation by geomagnetic field, which impacts the LEO proton and electron intensities. The short term variation in a single orbit could vary by a magnitude of several times. While the long term variation derived from the daily FSA rate over one year indicated the period of $\sim$ 57 days orbit precession with a magnitude change less than 20\% and a secular reduction at a rate of approximately $5\times 10^{-5}$\,counts/s/day. 

Distinct fluorescence lines from Al, Fe, Ni, and Cu are clearly observed in the instrumental background spectra. These lines could be employed for the in-orbit energy calibration. The spectral shape in the 0.5--10\,keV band remains consistent at different rate regimes. Furthermore, the obtained FSA rate displays a good linear correlation with the IMG, thanks to the innovative FF readout design of FXT. Based on these findings, the IMG instrumental background modelling was established, and validated by the FWC and the Lockman Hole observations. This modelling could facilitate the analysis of extended source observations such as those of galaxy clusters.

\begin{acknowledgments}
This work is based on data obtained with Einstein Probe, a space mission supported by Strategic Priority Program on Space Science of Chinese Academy of Sciences, in collaboration with ESA, MPE and CNES (Grant No. XDA15310303, No. XDA15310103, No. XDA15052100). 
\end{acknowledgments}

\bibliography{fxtbkg}{}

\begin{thebibliography}{}
\expandafter\ifx\csname natexlab\endcsname\relax\def\natexlab#1{#1}\fi
\providecommand{\url}[1]{\href{#1}{#1}}
\providecommand{\dodoi}[1]{doi:~\href{http://doi.org/#1}{\nolinkurl{#1}}}
\providecommand{\doeprint}[1]{\href{http://ascl.net/#1}{\nolinkurl{http://ascl.net/#1}}}
\providecommand{\doarXiv}[1]{\href{https://arxiv.org/abs/#1}{\nolinkurl{https://arxiv.org/abs/#1}}}

\bibitem[{E. {Bulbul} {et~al.}(2020){Bulbul}, {Kraft}, {Nulsen}, {Freyberg},
  {Miller}, {Grant}, {Bautz}, {Burrows}, {Allen}, {Eraerds}, {Fioretti},
  {Gastaldello}, {Ghirardini}, {Hall}, {Meidinger}, {Molendi}, {Rau},
  {Wilkins}, \& {Wilms}}]{2020ApJ...891...13B}
{Bulbul}, E., {Kraft}, R., {Nulsen}, P., {et~al.} 2020,
  \bibinfo{title}{{Characterization of the Particle-induced Background of
  XMM-Newton EPIC-pn: Short- and Long-term Variability},} \apj, 891, 13,
  \dodoi{10.3847/1538-4357/ab698a}

\bibitem[{J. {Cao} {et~al.}(2023){Cao}, {Chen}, {He}, {Xu}, {Chen}, {Gao},
  {Du}, {Wu}, \& {Liu}}]{2023OptEn..62d4103C}
{Cao}, J., {Chen}, T., {He}, H., {et~al.} 2023, \bibinfo{title}{{Visible light
  performance of x-ray filters for Einstein probe},} Optical Engineering, 62,
  044103, \dodoi{10.1117/1.OE.62.4.044103}

\bibitem[{Y. {Chen} {et~al.}(2025){Chen}, {Wang}, {Yang}, {Zhao}, {Ma}, {Wang},
  {Han}, {Cui}, {Ban}, {Bi}, {Bianucci}, {Burwitz}, {Cai}, {Cai}, {Chen},
  {Chen}, {Chen}, {Cheng}, {Ding}, {Eder}, {Feng}, {Feng}, {Friedrich}, {Geng},
  {Gao}, {Gao}, {Guan}, {Hou}, {Huo}, {Jia}, {Keereman}, {Li}, {Li}, {Li},
  {Li}, {Li}, {Li}, {Li}, {Liao}, {Liu}, {Liu}, {Liu}, {Lu}, {Luo},
  {Meidinger}, {Qiang}, {Santovincenzo}, {Sheng}, {Sun}, {Tang}, {Tian},
  {Valsecchi}, {Vernani}, {Wang}, {Wang}, {Wang}, {Wang}, {Wang}, {Wang},
  {Xie}, {Xu}, {Xue}, {Yan}, {Yang}, {Yu}, {Yuan}, {Zhang}, {Zhang}, {Zhang},
  {Zhang}, {Zhang}, {Zhang}, {Zhang}, {Zhao}, {Zhao}, \&
  {Zhu}}]{2025RDTM..tmp...65C}
{Chen}, Y., {Wang}, J., {Yang}, Y., {et~al.} 2025, \bibinfo{title}{{Design and
  development of the follow-up X-ray telescope onboard Einstein Probe in China:
  a review},} Radiation Detection Technology and Methods,
  \dodoi{10.1007/s41605-025-00558-0}

\bibitem[{H. {Cheng} {et~al.}(2024){Cheng}, {Ling}, {Zhang}, {Sun}, {Sun},
  {Liu}, {Dai}, {Jia}, {Pan}, {Wang}, {Zhao}, {Chen}, {Cheng}, {Fu}, {Han},
  {Li}, {Li}, {Ma}, {Xue}, {Yan}, {Zhang}, {Wang}, {Yang}, {Zhao}, \&
  {Yuan}}]{2024ExA....57...10C}
{Cheng}, H., {Ling}, Z., {Zhang}, C., {et~al.} 2024, \bibinfo{title}{{Ground
  calibration result of the Lobster Eye Imager for Astronomy},} Experimental
  Astronomy, 57, 10, \dodoi{10.1007/s10686-024-09932-0}

\bibitem[{W. {Cui} {et~al.}(2023){Cui}, {Wang}, {Zhao}, {Zhang}, {Meidinger},
  {Yang}, {Keil}, {Zhang}, {Huo}, {Wang}, {Song}, {Lu}, {Ma}, {Wang}, {Xu},
  {Zhu}, {Li}, {Li}, {Luo}, {Han}, {Zhao}, {Hou}, {Yang}, {Geng}, {Li}, {Chen},
  {Tang}, {Chen}, \& {Chen}}]{2023ExA....55..603C}
{Cui}, W., {Wang}, H., {Zhao}, X., {et~al.} 2023, \bibinfo{title}{{Design and
  performance of the focal plane camera for FXT onboard the Einstein Probe
  satellite},} Experimental Astronomy, 55, 603,
  \dodoi{10.1007/s10686-023-09891-y}

\bibitem[{M. {Freyberg} {et~al.}(2021){Freyberg}, {Perinati}, {Pacaud},
  {Eraerds}, {Churazov}, {Dennerl}, {Predehl}, {Merloni}, {Meidinger},
  {Bulbul}, {Friedrich}, {Gilfanov}, {Tenzer}, {Pommranz}, {Eckert}, {Schmitt},
  {Brusa}, \& {Santangelo}}]{2021SPIE11444E..1OF}
{Freyberg}, M., {Perinati}, E., {Pacaud}, F., {et~al.} 2021,
  \bibinfo{title}{{SRG/eROSITA in-flight background at L2},} in Society of
  Photo-Optical Instrumentation Engineers (SPIE) Conference Series, Vol. 11444,
  Society of Photo-Optical Instrumentation Engineers (SPIE) Conference Series,
  ed. J.-W.~A. {den Herder}, S.~{Nikzad}, \& K.~{Nakazawa}, 114441O,
  \dodoi{10.1117/12.2562709}

\bibitem[{M. {Gerontidou} {et~al.}(2021){Gerontidou}, {Katzourakis},
  {Mavromichalaki}, {Yanke}, \& {Eroshenko}}]{2021AdSpR..67.2231G}
{Gerontidou}, M., {Katzourakis}, N., {Mavromichalaki}, H., {Yanke}, V., \&
  {Eroshenko}, E. 2021, \bibinfo{title}{{World grid of cosmic ray vertical
  cut-off rigidity for the last decade},} Advances in Space Research, 67, 2231,
  \dodoi{10.1016/j.asr.2021.01.011}

\bibitem[{C.~E. {Grant} {et~al.}(2022){Grant}, {Miller}, {Bautz}, {Foster},
  {Kraft}, {Allen}, \& {Burrows}}]{2022SPIE12181E..2EG}
{Grant}, C.~E., {Miller}, E.~D., {Bautz}, M.~W., {et~al.} 2022,
  \bibinfo{title}{{Towards precision particle background estimation for future
  x-ray missions: correlated variability between Chandra ACIS and AMS},} in
  Society of Photo-Optical Instrumentation Engineers (SPIE) Conference Series,
  Vol. 12181, Space Telescopes and Instrumentation 2022: Ultraviolet to Gamma
  Ray, ed. J.-W.~A. {den Herder}, S.~{Nikzad}, \& K.~{Nakazawa}, 121812E,
  \dodoi{10.1117/12.2629520}

\bibitem[{H. {Koshiishi}(2014){Koshiishi}}]{2014AdSpR..53..233K}
{Koshiishi}, H. 2014, \bibinfo{title}{{Space radiation environment in low earth
  orbit during influences from solar and geomagnetic events in December 2006},}
  Advances in Space Research, 53, 233, \dodoi{10.1016/j.asr.2013.11.014}

\bibitem[{J.-Y. {Liao} {et~al.}(2020){Liao}, {Zhang}, {Chen}, {Zhang}, {Jin},
  {Chang}, {Chen}, {Ge}, {Guo}, {Li}, {Li}, {Lu}, {Lu}, {Nie}, {Song}, {Yang},
  {You}, {Zhao}, \& {Zhang}}]{2020JHEAp..27...24L}
{Liao}, J.-Y., {Zhang}, S., {Chen}, Y., {et~al.} 2020,
  \bibinfo{title}{{Background model for the Low-Energy Telescope of
  Insight-HXMT},} Journal of High Energy Astrophysics, 27, 24,
  \dodoi{10.1016/j.jheap.2020.02.010}

\bibitem[{P. {Predehl} {et~al.}(2021){Predehl}, {Andritschke}, {Arefiev},
  {Babyshkin}, {Batanov}, {Becker}, {B{\"o}hringer}, {Bogomolov}, {Boller},
  {Borm}, {Bornemann}, {Br{\"a}uninger}, {Br{\"u}ggen}, {Brunner}, {Brusa},
  {Bulbul}, {Buntov}, {Burwitz}, {Burkert}, {Clerc}, {Churazov}, {Coutinho},
  {Dauser}, {Dennerl}, {Doroshenko}, {Eder}, {Emberger}, {Eraerds},
  {Finoguenov}, {Freyberg}, {Friedrich}, {Friedrich}, {F{\"u}rmetz},
  {Georgakakis}, {Gilfanov}, {Granato}, {Grossberger}, {Gueguen}, {Gureev},
  {Haberl}, {H{\"a}lker}, {Hartner}, {Hasinger}, {Huber}, {Ji}, {Kienlin},
  {Kink}, {Korotkov}, {Kreykenbohm}, {Lamer}, {Lomakin}, {Lapshov}, {Liu},
  {Maitra}, {Meidinger}, {Menz}, {Merloni}, {Mernik}, {Mican}, {Mohr},
  {M{\"u}ller}, {Nandra}, {Nazarov}, {Pacaud}, {Pavlinsky}, {Perinati},
  {Pfeffermann}, {Pietschner}, {Ramos-Ceja}, {Rau}, {Reiffers}, {Reiprich},
  {Robrade}, {Salvato}, {Sanders}, {Santangelo}, {Sasaki}, {Scheuerle},
  {Schmid}, {Schmitt}, {Schwope}, {Shirshakov}, {Steinmetz}, {Stewart},
  {Str{\"u}der}, {Sunyaev}, {Tenzer}, {Tiedemann}, {Tr{\"u}mper}, {Voron},
  {Weber}, {Wilms}, \& {Yaroshenko}}]{2021A&A...647A...1P}
{Predehl}, P., {Andritschke}, R., {Arefiev}, V., {et~al.} 2021,
  \bibinfo{title}{{The eROSITA X-ray telescope on SRG},} \aap, 647, A1,
  \dodoi{10.1051/0004-6361/202039313}

\bibitem[{D.~F. {Smart} \& M.~A. {Shea}(2009){Smart} \&
  {Shea}}]{2009AdSpR..44.1107S}
{Smart}, D.~F., \& {Shea}, M.~A. 2009, \bibinfo{title}{{Fifty years of progress
  in geomagnetic cutoff rigidity determinations},} Advances in Space Research,
  44, 1107, \dodoi{10.1016/j.asr.2009.07.005}

\bibitem[{H. {Suzuki} {et~al.}(2021){Suzuki}, {Plucinsky}, {Gaetz}, \&
  {Bamba}}]{2021A&A...655A.116S}
{Suzuki}, H., {Plucinsky}, P.~P., {Gaetz}, T.~J., \& {Bamba}, A. 2021,
  \bibinfo{title}{{Spatial and temporal variations of the Chandra ACIS
  particle-induced background and development of a spectral-model generation
  tool},} \aap, 655, A116, \dodoi{10.1051/0004-6361/202141458}

\bibitem[{Y. {Yang} {et~al.}(2023){Yang}, {Wang}, {Han}, {Wang}, {Cui}, {Zhu},
  {Cong}, {Ma}, {Zhao}, {Hou}, {Yang}, {Chen}, {Lu}, {Lv}, {Sun}, {Zhang},
  {Yu}, {Wang}, {Liu}, {Zhang}, {Bi}, {Lu}, {Friedrich}, {Eder}, {Hartmann},
  {Burwitz}, {Keereman}, {Santovincenzo}, {Vernani}, {Bianucci}, {Valsecch},
  {Sheng}, {Yan}, {Qiang}, {Wang}, {Wang}, {Wang}, {Ding}, {Wang}, {Cheng}, \&
  {Chen}}]{2023ExA....55..625Y}
{Yang}, Y., {Wang}, Y., {Han}, D., {et~al.} 2023, \bibinfo{title}{{Design and
  testing of the Optics for FXT onboard EP satellite},} Experimental Astronomy,
  55, 625, \dodoi{10.1007/s10686-022-09870-9}

\bibitem[{W. {Yuan} {et~al.}(2022){Yuan}, {Zhang}, {Chen}, \&
  {Ling}}]{2022hxga.book...86Y}
{Yuan}, W., {Zhang}, C., {Chen}, Y., \& {Ling}, Z. 2022, \bibinfo{title}{{The
  Einstein Probe Mission},} in Handbook of X-ray and Gamma-ray Astrophysics,
  ed. C.~{Bambi} \& A.~{Sangangelo}, 86,
  \dodoi{10.1007/978-981-16-4544-0_151-1}

\bibitem[{W. {Yuan} {et~al.}(2016){Yuan}, {Amati}, {Cannizzo}, {Cordier},
  {Gehrels}, {Ghirlanda}, {G{\"o}tz}, {Produit}, {Qiu}, {Sun}, {Tanvir}, {Wei},
  \& {Zhang}}]{2016SSRv..202..235Y}
{Yuan}, W., {Amati}, L., {Cannizzo}, J.~K., {et~al.} 2016,
  \bibinfo{title}{{Perspectives on Gamma-Ray Burst Physics and Cosmology with
  Next Generation Facilities},} \ssr, 202, 235,
  \dodoi{10.1007/s11214-016-0274-z}

\bibitem[{W. {Yuan} {et~al.}(2018{\natexlab{a}}){Yuan}, {Zhang}, {Chen}, {Sun},
  {Zhang}, {Cui}, {Ling}, {Huang}, {Zhao}, {Wang}, {Qiu}, {Liu}, {Pan}, {Cai},
  {Deng}, {Jia}, {Jin}, {Sun}, {Hu}, {Liu}, {Zhang}, {Song}, {Lu}, {Jia}, {Li},
  {Zhao}, {Ge}, {Zhang}, {Cui}, {Wang}, {Wang}, {Sun}, {Jin}, {Li}, {Chen},
  {Cai}, {Guo}, {Liu}, {Liu}, {Feng}, {Zhang}, {Zhang}, {Dai}, {Wu}, \&
  {Gou}}]{2018SSPMA..48c9502Y}
{Yuan}, W., {Zhang}, C., {Chen}, Y., {et~al.} 2018{\natexlab{a}},
  \bibinfo{title}{{Einstein Probe: Exploring the ever-changing X-ray
  Universe},} Scientia Sinica Physica, Mechanica \& Astronomica, 48, 039502,
  \dodoi{10.1360/SSPMA2017-00297}

\bibitem[{W. {Yuan} {et~al.}(2018{\natexlab{b}}){Yuan}, {Zhang}, {Ling},
  {Zhao}, {Wang}, {Chen}, {Lu}, {Zhang}, \& {Cui}}]{2018SPIE10699E..25Y}
{Yuan}, W., {Zhang}, C., {Ling}, Z., {et~al.} 2018{\natexlab{b}},
  \bibinfo{title}{{Einstein Probe: a lobster-eye telescope for monitoring the
  x-ray sky},} in Society of Photo-Optical Instrumentation Engineers (SPIE)
  Conference Series, Vol. 10699, Space Telescopes and Instrumentation 2018:
  Ultraviolet to Gamma Ray, ed. J.-W.~A. {den Herder}, S.~{Nikzad}, \&
  K.~{Nakazawa}, 1069925, \dodoi{10.1117/12.2313358}

\bibitem[{J. {Zhang} {et~al.}(2022){Zhang}, {Qi}, {Yang}, {Wang}, {Liu}, {Cui},
  {Zhao}, {Jia}, {Li}, {Chen}, {Li}, {Zhao}, {Chen}, {Liu}, {Bao}, {Guan},
  {Song}, \& {Yuan}}]{2022APh...13702668Z}
{Zhang}, J., {Qi}, L., {Yang}, Y., {et~al.} 2022, \bibinfo{title}{{Estimate of
  the background and sensitivity of the follow-up X-ray telescope onboard
  Einstein Probe},} Astroparticle Physics, 137, 102668,
  \dodoi{10.1016/j.astropartphys.2021.102668}

\bibitem[{D. {Zhao} {et~al.}(2017){Zhao}, {Zhang}, {Yuan}, {Zhang},
  {Willingale}, \& {Ling}}]{2017ExA....43..267Z}
{Zhao}, D., {Zhang}, C., {Yuan}, W., {et~al.} 2017, \bibinfo{title}{{Geant4
  simulations of a wide-angle x-ray focusing telescope},} Experimental
  Astronomy, 43, 267, \dodoi{10.1007/s10686-017-9534-5}

\bibitem[{X. {Zhao} {et~al.}(2025){Zhao}, {Cui}, {Wang}, {Zhang}, {Zhao},
  {Hou}, {Zhu}, {Wang}, {Xu}, {Luo}, {Han}, {Yang}, {Wang}, {Ma}, {Yang},
  {Huo}, {Li}, {Zhang}, {Geng}, \& {Chen}}]{2025RAA....25a5002Z}
{Zhao}, X., {Cui}, W., {Wang}, H., {et~al.} 2025, \bibinfo{title}{{Timing
  Calibration of the Follow-up X-Ray Telescope On Board the Einstein Probe
  Satellite},} Research in Astronomy and Astrophysics, 25, 015002,
  \dodoi{10.1088/1674-4527/ad981b}

\end{thebibliography}
\bibliographystyle{aasjournalv7}


\end{CJK*}
\end{document}